\documentclass[%
 reprint,
 amsmath,amssymb,
 aps,
]{revtex4-2}
\usepackage{natbib}
\usepackage[colorlinks=true,linkcolor= blue
,citecolor=blue]{hyperref}%

\newcommand{\bk}{{\bf k}}

\newcommand{\bp}{{\bf p}}

\newcommand{\br}{{\bf r}}

\newcommand{\bR}{{\bf R}}
\newcommand{\bP}{{\bf P}}

\newcommand{\beq}{\begin{equation}}
\newcommand{\beqn}{\begin{eqnarray}}
\newcommand{\eeq}{\end{equation}}
\newcommand{\eeqn}{\end{eqnarray}}

\usepackage{amsmath,amsfonts,amssymb}

\usepackage{appendix}
\usepackage{graphicx}
\usepackage{dcolumn}
\usepackage{bm,color}

\begin{document}


\title{ Berry curvature contribution towards $ 1s-2p_{\pm} $  interlayer exciton   ultrafast transition within a $R-WSe_{2}/MoSe_{2} $ heterobilayer. }

\author{H. E. Hannachi}
\email{hannachi.houssemedine@fsb.ucar.tn}

 \affiliation{Laboratoire de Physique des Mat\'eriaux, Facult\'e des Sciences de Bizerte, Universit\'e de Carthage, 7021 Zarzouna, Tunisie.}

\author{M. O. Goerbig}
\email{mark-oliver.goerbig@universite-paris-saclay.fr}
\affiliation{Laboratoire de Physique des Solides, CNRS UMR 8502, Universit\'e Paris-Saclay, 91405 Orsay Cedex, France}

\author{ S. Jaziri}
 \email{Sihem.Jaziri@fsb.rnu.tn}
  \affiliation{Laboratoire de Physique des Mat\'eriaux, Facult\'e des Sciences de Bizerte, Universit\'e de Carthage, 7021 Zarzouna, Tunisie.}
\affiliation{Laboratoire de Physique de la Mati\`ere Condens\'ee, Facult\'e des Sciences de Tunis, Universit\'e de Tunis El Manar, 2092 El Manar, Tunisie}%

\date{\today}

\begin{abstract}

We calculate the spectrum of interlayer neutral excitons in transition-metal-dichalcogenide WeSe$_2$/MoSe$_2$ heterobilayers in the $R$-stacking configuration. Most saliently, we show that, similarly to neutral excitons and trions in monolayer transition-metal dichalcogenides, the spectrum is sensitive to the Berry curvature and thus quantum-geometric effects underlying the electron and hole wave functions. Due to the spatial separation between the electron and hole constituting the exciton in different layers, the Berry-curvature-induced splitting of the between the $2p_+$ and the $2p_-$ exciton states is smaller than for monolayer excitons. Furthermore, we investigate the dependence of the exciton spectra on the dielectric environment and the twist angle between the two layers. Finally, the long-lived moiré interlayer exciton ground state ($1s$) enhances the possibility of creating brightened $2p_{\pm}$ states using a circularly polarized medium-infrared probe from the $1s$ ground state. As a result, we determine the polarizability of the $1s-2p_{\pm}$ transition, following by two-level dressed model for the optical Stark effect.

\end{abstract}

\maketitle


\section{Introduction}

The interaction between ultrafast lasers and two-dimensional (2D) semiconductors has attracted significant attention within the field of ultrafast optoelectronics \cite{Pettine2023, Long2019}. This interaction does not only offer an effective means of characterizing band structures but also facilitates nonlinear optical responses \cite{Peres2021,Shree2021}. Recently, the exploration of transition-metal-dichalcogenide (TMD) bilayers has provided an exceptional platform for studying ultrafast emission phenomena, crucial for the development of high-performance speed devices \cite{Kim2021, merkl}. 

From a more theoretical point of view, the intriguing quantum phenomena are intricately linked to the geometric and topological characteristics of the low-energy charge carriers in 2D semiconducting TMD systems \cite{donald2017,Hu2023}. Indeed, the direct gap is situated at the $K$ and $K'$ points in reciprocal space -- the corners of the first Brillouin zone -- which are not time-reversal-invariant momenta, where the Berry curvature as the most important quantum-geometric quantity would need to vanish. Indeed, in 2D TMD semiconductors, the Berry curvature is even maximal at these points. 

The Berry curvature, which may be viewed as a reciprocal-space pseudomagnetic field, has significantly contributed to the emergence of exotic transport phenomena in TMD systems, including the anomalous Hall effect \cite{Xie2022}, spin-valley Hall effect \cite{Lee2021}, and more recently, the nonlinear Hall effect \cite{Hu20}. Furthermore, it has unveiled several optical phenomena, such as the fine structure of excitons and their non-hydrogenic spectra in TMD monolayers  \cite{aida,trushin0,trushin,vastana,zhou,Wu2015,Chang2021,Tang2023}. Specifically, the spontaneous energy splitting between $2p_{+}$ and $2p_{-}$ dark states in MoSe$_{2}$ monolayer, predicted to be between 10-20 meV, has been identified as a Berry curvature effect on the exciton spectrum. This observation has recently been confirmed through experimental evidence utilizing optical Stark  spectroscopy \cite{Chaw}. 

Beyond monolayer TMD systems, photoluminescence spectroscopy experiments have demonstrated that the optical response of the TMDs heterobilayers is predominantly influenced by both intralayer and interlayer excitons \cite{tran,evgeny,devakul,pasqual}. These two types of excitons are identified based on whether the electron and hole reside within the same or different layers constituting the bilayer system. Owing to the spatial separation in bilayer systems, the interlayer neutral exciton (IX) is generically associated with a lower energy scale and exhibits a weaker light-matter coupling than their intralayer counterpart \cite{donald2018,hichri,Houssem2023,david}. In contrast, they demonstrate an extraordinarily long lifetime, extending up to 100 ns, surpassing that of intralayer excitons \cite{Houssem2023,rivera,xiabo,lifetime}. Consequently, IXs are regarded as promising candidates for the phenomena of excitons condensation \cite{Wang2019,Gua2022}.  

Contrary to monolayer systems, the effect of the Berry phase on interlayer excitons in TMDs heterobilayers has, to the best of our knowledge, not been fully studied so far. This is the main aim of the present paper. Most interestingly, the separation between the two layers and the sensitivity of the center-of-mass and relative coordinates to the moiré potential induced by a twist or stacking mismatch between the two layers make the bilayer system considerably more complex as compared to their TMD monolayer counterpart. However, recent theoretical works have started to shed some light on this phenomenon. Very recent theoretical studies discuss the influence of the Berry phase on localized excitons in bilayer systems \cite{Tang2023, Knapp2021}. These studies show that the Berry phase manifests itself as an anomalous Hall velocity of the interlayer exciton's center-of-mass coordinate when two external in-plane electric fields are applied to the electron and hole constituents, respectively. Furthermore, another study investigated the electron and hole Berry curvatures residing in different layers for $R$ and $H$ bilayer stackings \cite{Kormanyos2018}. It was shown that in $R$ stacking, the electron and hole Berry curvatures have opposite signs, while in $H$ stacking, they have the same sign. This implies that the Berry curvature affects the interlayer-exciton spectra in $R$ stacking much more significantly than for $H$-stacked bilayers, as the exciton Berry curvature is determined by the difference between the electron and hole Berry curvatures \cite{aida,trushin,vastana,zhou}. 

In the present paper, we concentrate our study on the quasi-direct interlayer exciton case forming around the $ K_{m}(K_{m}^{'})$ valley points at weak twist angles with respect to  a $R$-stacked WSe$_{2}$/MoSe$_{2}$ heterobilayer. We calculate the influence of the Berry curvature on the IX spectra, taking into account several experimentally tunable parameters such as the interlayer distance, the dielectric environment and the twist angle. Furthermore, we present a model that describes the brightening of the $2p_{\pm}$ dark state by resonantly probing the $1s-2p_{\pm}$ interlayer exciton transition. Given the long-lived interlayer ground state, this transition can be observed using near-infrared (NIR)-pump/medium-infrared (MIR)-probe spectroscopy, as that used by Merkl \textit{et al.} \cite{merkl}. We anticipate that the Berry curvature effects on the relative exciton motion could induce the $1s-2p_{\pm}$ transition with two degrees of freedom that can be probed by optical circular polarization. This is of particular interest for high-performance, high-speed devices. 

This paper is organized as follows. In Sec. \ref{sec2}, we provide a self-contained introduction to the moiré interlayer exciton model, with a specific focus on the influence of Berry curvature on the relative motion of interlayer excitons using the Mott-Wannier exciton equation. Section \ref{sec3} is devoted to our numerical results. In a first step, we determine the binding energies and corresponding eigenvectors for both interlayer and intralayer excitons, taking into consideration the Berry curvature. The spectra are then discussed when varying several experimentally relevant factors, including the dielectric environment, interlayer distance and twisting effects. This allows us to assess the impact of Berry curvature on the binding energies and splitting of the pertinent IX $2p$ levels. Finally, our attention is directed towards the interlayer exciton polarizability and the influence of the Berry curvature on the $1s-2p_{\pm}$ exciton transition. Our approach involves presenting a model that examines the optical Stark effect of a two-level system. Our conclusions may be found in Sec. \ref{sec5}.

 \section{Theoretical framework} \label{sec2}
 
\begin{figure}
\begin{center}
 \includegraphics[scale=0.6]{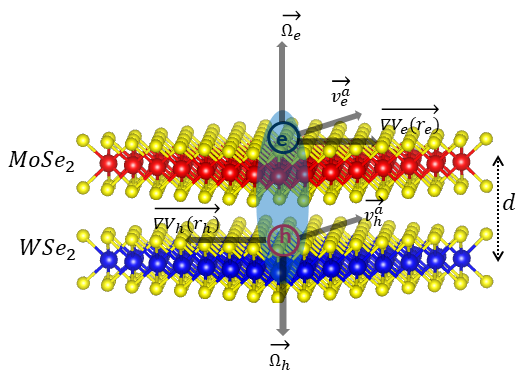}
 \caption{  Illustration of  interlayer exciton formation within the $ WSe_{2}/MoSe_{2} $ heterobilayer, where d represents the interlayer distance.  The electron and hole each possess  opposing Berry curvature vectors,  denoted as $ \boldsymbol{\Omega}_{e} $ and $\boldsymbol{\Omega}_{h} $, respectively.  The electron and hole anomalous velocity $ \textbf{v}_{e}^{a} $ and $ \textbf{v}_{h}^{a}$, align in the same direction. Additionally, the third vector $\nabla V_{e(h)}(\textbf{r}_{e(h)}) $ illustrates the electric force acting on the electron (hole) due to the local potential at its position, $ \textbf{r}_{e(h)} $.}\label{fig1}
  \end{center}
\end{figure}

 We consider the formation of an interlayer exciton  within a $R-WSe_{2}/MoSe_{2} $ van der Waals heterobilayer, where the electron resides in the minimum of the conduction band of the $MoSe_{2}$ monolayer and the hole occupies the maximum of the valence band in the $WSe_{2}$ monolayer. The charge carriers are characterized by their respective band masses $m_{e(h)}$ and their momenta $\textbf{p}_{e(h)}=\hbar \textbf{k}_{e(h)}$, where the subscript indicates either an electron in the conduction band, $e$, or a hole in the valence band, $h$. Since the extrema of the two bands, where excitons are formed, are situated at the $K_{m}$ and $K'_{m}$ points of the first Brillouin zone and optically coupled, they are associated with a non-zero Berry curvature, $ \boldsymbol{\Omega}_{e(h)}(\textbf{k}_{e(h)}) =\nabla_{\textbf{k}_{e(h)}}\times \textbf{A}_{e(h)}(\textbf{k}_{e(h)})$, where $\textbf{A}_{e(h)}(\textbf{k}_{e(h)})=i\langle U_{\textbf{k}_{e(h)}}(\textbf{r}_{e(h)})|\nabla_{\textbf{k}_{e(h)}}|U_{\textbf{k}_{e(h)}}(\textbf{r}_{e(h)})\rangle$ is the Berry connection in terms of the Bloch states $ U_{\textbf{k}_{e(h)}}(\textbf{r}_{e(h)})$, which are associated with the periodic part  of the Bloch wave function. The presence  of a non-zero Berry curvature and a local potential $ V_{e(h)}(\textbf{r}_{e(h)}) $, both lead to an electron(hole) anomalous velocity $ \textbf{v}_{e(h)}^{a} $, that is perpendicular to the corresponding Berry curvature vector and to the associated local force (see figure \ref{fig1}). Thus, the dynamical properties are governed by the following  semi-classical equations  motions, in the absence of a magnetic field,
 \begin{eqnarray}
 \frac{d \textbf{p}_{e(h)}}{dt} &=& -\nabla_{r_{e(h)}}V_{e(h)}(r_{e(h)})\qquad \text{and}
 \\ 
\frac{d\textbf{r}_{e(h)}}{dt} &=& \frac{1}{\hbar}\nabla_{\textbf{k}_{e(h)}}\xi^{e(h)}(\textbf{p}_{e(h)})-\frac{d \textbf{k}_{e(h)}}{dt}\times \boldsymbol{\Omega}_{e(h)}.
\end{eqnarray}
The first equation represents the applied force on the charges. The second one corresponds to the velocity of the electron or hole, respectively, and has two terms. The first term  $\boldsymbol{\upsilon}^{e(h)}_{g}=\nabla_\bk\xi^{e(h)}(\textbf{k}_{e(h)})/\hbar$ is the usual group velocity derived from the dispersion relation 
\begin{equation}
    \xi^{e(h)}(\textbf{k}_{e(h)})=\pm \sqrt{\triangle^{2}_{Mo(W)}+\hbar v^{2}_{Mo(W)}\textbf{k}^{2}_{e(h)}} 
\end{equation}
where $ \triangle_{Mo(W)} $ is the half gap for $ MoSe_{2} $($ WSe_{2} $) monolayers. The velocity parameter $\alpha_{Mo(W)}\propto (at)_{Mo(W)}/\hbar$ is given in terms of the lattice spacing $a$ and the characteristic hopping integral $t$, in a tight-binding description. The second term $ \textbf{v}_{e(h)}^{a}=- (d \textbf{k}_{e(h)}/dt)\times \boldsymbol{\Omega}_{e(h)}$ is the electron(hole) anomalous velocity. This component modifies the electron (hole) quantum Hamiltonian and  gives rise to a non-commutativity between the position and momentum operators \cite{aida,trushin,vastana,zhou}. In addition to the anomalous velocity, this term happens to play a role when Bloch electrons are exposed to an electric field that is not that generated by the periodic lattice potential but varying slowly at the lattice scale. As we discuss in more detail below, the mutual electric potential due to the electrostatic interaction between the electron and the hole forming the exciton also affects the excitons' spectral properties via the Berry curvature.


\subsection{Moir\'e potential}

\begin{figure*}
\begin{center}
 \includegraphics[scale=0.6]{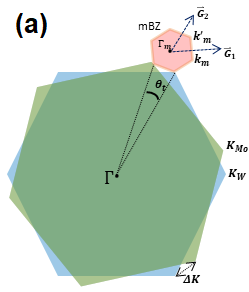}
 \quad
 \includegraphics[scale=0.59]{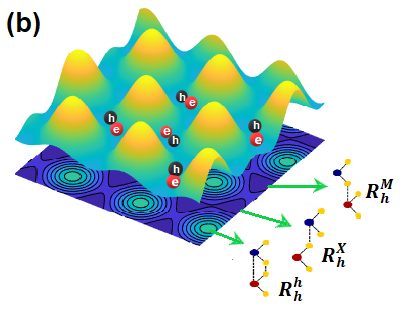}
 \quad
 \includegraphics[scale=0.48]{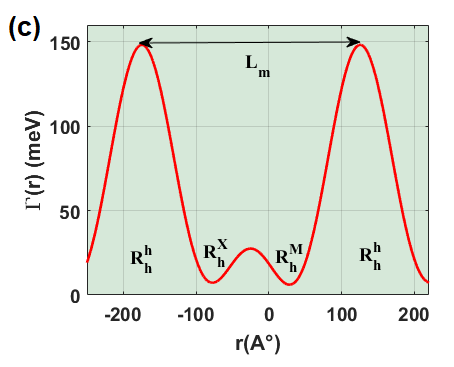}
 \caption{ Figure (a) illustrates the formation of moiré mini Brillouin zone (mBZ) of twisted $MoSe_2/WSe_2$ heterobilayer.
 Figure (b) depicts the 2D periodic moiré potential, portrayed in both 3D graph and 2D projection, effectively confining the interlayer excitons (represented by red and black spheres) within localized minima. This visualization eloquently highlights the intricate moiré patterns denoted as $R_{h}^{h}$, $R_{h}^{X}$, and $R_{h}^{M}$. In figure (c), a lateral profile of $\Gamma(\textbf{r})$ at a 1° twist angle is presented, meticulously delineating the moiré period from the $R_{h}^{h}$ to $R_{h}^{h}$ pattern.  }\label{potential}
  \end{center}
\end{figure*}

In addition to the intrinsic properties of the excitons in heterobilayers, the latter are also affected by a possible stacking mismatch of the TMD layers. In our case, when $WSe_{2}$ and $MoSe_{2}$ are stacked at a weak twist angle $\theta_{t}$, a valley mismatch arises due to the rotational mismatch of the associated Brillouin zones. This mismatch is characterized by the vector $\boldsymbol{\Delta K}= \textbf{K}_{Mo}-\textbf{K}_{W}$, which measures the reciprocal-space distance between equivalent $K$ points associated with the two layers. Additionally, moiré patterns emerge, exhibiting a long moiré period $L_{m}>100$ nm that characterizes a new unit cell much larger than those in the individual layers. In reciprocal space, this larger unit cell yields a mini Brillouin zone (mBZ) featuring novel high-symmetry points denoted as $K_m$, $\Gamma_m$  and $K'_m$, as depicted in figure \ref{potential}(a). For a weak twist angle, the moiré period is approximated as
$$ L_{m}=\frac{a_{0<}}{\sqrt{\theta^{2}_{t}+\delta_{0}^{2}}} ,$$ 
where $ \delta_{0}=1-a_{0<}/a_{0>} $ is lattice mismatch and $ a_{0<(>)} $ is the inferior  (superior) lattice constant of the TMD monolayers. The moiré superlattice is composed of three distinct atomic registries  denoted by $ R^{h}_{h} $, $ R^{X}_{h} $, and $ R^{M}_{h} $, interconnected by the $ C_{3} $ group symmetry \cite{tran}. Here, $R^{h}_{h}$ represents a hexagonal registry where the transition metal atoms of the $WSe_2$ layer align directly on top of the transition metal atoms of the $MoSe_2$ layer. In contrast, $R^{X}_{h}$ and $R^{M}_{h}$ stackings are characterized by a lateral shift between the two layers, where the transition metal atoms of one layer align directly with the chalcogen atoms of the other layer. These two atomic configurations exhibit $C^{1}_{3}$ and $C^{2}_{3}$ rotation symmetries with respect to the $R^{h}_{h}$ stacking, respectively [see Fig. \ref{potential}(b)].
This moiré network can be described by a periodic moir\'e potential \cite{donald2018,tran}  $ \Gamma (\textbf{r})=2V_{0}\sum_{j=1,2,3} \exp(i\textbf{G}_{j}\textbf{r})$ that respects the underlying $C_3$ symmetry of the moir\'e lattice [see figures \ref{potential}(b) and (c)]. Here $V_{0}\sim 9$\,meV is the amplitude of the modulation the value of which has been determined by DFT calculations \cite{tran}.  Notice that this potential is not an electrostatic one that acts differently on the two types of charges but it represents a spatial modulation of the minimal gap between conduction and valence band. Both electrons and holes are therefore attracted by the minima of this potential so that it can be directly viewed as an exciton potential. The reciprocal lattice vectors $ \textbf{G}_{j} $ are associated with the moiré supercell and can be obtained from those, $ \textbf{b}_{j} $, of the (untwisted) monolayer and the twist angle $\theta_t$ between the two layers, $ \textbf{G}_{j}=\theta_{t}\textbf{b}_{j}\times \textbf{e}_{z} $. The $ \textbf{G}_{j} $ vectors are given by $ \textbf{G}_{1}=(4\pi/\sqrt{3}L_{m})(1,0) $, $ \textbf{G}_{2}=C^{1}_{3}\textbf{G}_{1} $ and $ \textbf{G}_{3}=C^{2}_{3}\textbf{G}_{1} $, in terms of the rotations $C_3^j$ of the $C_3$ group.  Near the minima of the moiré potential [as shown in the Figs. \ref{potential}(b) and \ref{potential}(c)], 
we might consider the latter to be parabolic within a series expansion, 
$ V_{M}(\textbf{r}_{\alpha})= m_{\alpha}\omega_{\alpha}^{2}\textbf{r}_{\alpha}^{2}/2$ \cite{tran}. Here, the subscript $\alpha = e, h$ designates the electron band as 'e' and the hole band as 'h', respectively. The term $\omega_{\alpha}$ signifies the frequency of charge carriers, while their corresponding moiré potential size is denoted by $R_{\alpha} = \sqrt{\hbar/m_{\alpha}\omega_{\alpha}}$. The frequency of charge carriers is anticipated to be expressed in terms of the moiré period as: $\omega_{\alpha} = 4\pi\sqrt{V_0/m_{\alpha}}/L_m$. Consequently, we derive $R_{\alpha} = \sqrt{\ell_{\alpha} L_{m}}$, where the characteristic length $\ell_{\alpha}$ is defined as $\ell_{\alpha} = \hbar/ 4\pi\sqrt{m_{\alpha}V_0}$. The estimated values for the characteristic lengths $\ell_e$ and $\ell_h$ are 2.58 Å and 3.45 Å, respectively, based on charge carriers' masses $m_e = 0.8$ and $m_h = 0.45$. As we show below, this approximation is justified since the moiré period is much larger than our other characteristic length scales of the exciton, namely the Bohr radius $a_B=\hbar^2\epsilon/e^2 \mu$ and the Compton length (the square root of the Berry curvature) $\lambda_C\sim\sqrt{\Omega}\sim\hbar /\sqrt{\mu\triangle}$, in terms of the reduced mass $ \mu=m_{e}m_{h}/M=m_em_h/(m_e+m_h) $. Both are on the order of some nm.

\subsection{Electron-hole interactions}

The Coulomb-type electron-hole interactions in the van-der-Waals heterostructure need to take into account non-local screening due to possible excitations from the valence to the conduction band over the effective gap $\Delta$. Notice that in the MoSe$_2$/WeSe$_2$ heterobilayer this gap is the energy difference between the top of the WeSe$_2$ (hole) valence band and the bottom of the MoSe$_2$ (electron) conduction band. To leading order in $q$, these excitations yield a dielectric function 
\begin{equation}
    \epsilon(q)\simeq 1+ \xi q,
\end{equation}
where the characteristic length
\begin{equation}\label{eq:chi}
    \xi=\#\frac{e^2}{\epsilon\Delta}
\end{equation}
is the polarizability of the 2D material embedded in a dielectric environment characterized by the dielectric constant $\epsilon$, where $\#$ is a numerical prefactor \cite{Marino}.  This polarizability yield precisely the screening length 
\begin{equation}\label{eq:scrl}
    \ell_s=2\pi \xi,
\end{equation}
which enters into the expression of the Rytova-Keldysh potential \cite{keldysh}
\begin{equation}\label{eq:eq2}
V_{RK}(r=|\br|)=-\pi R_y \frac{a_B}{\ell_s} K_{0}\left(\dfrac{\sqrt{r^{2}+d^{2}}}{\ell_{s}}\right),
\end{equation}
which is the 2D Fourier transform of the screened Coulomb interaction $V(q)=V_0(q)/\epsilon(q)$, where $V_0=2\pi e^2\exp(-qd)/q$ represents the bare Coulomb potential that takes into account the layer separation $d$, which is the minimal distance between the electron in the MoSe$_2$ layer and the hole in the WSe$_2$ layer.  Moreover, $|\br|=\br_e-\br_h$ is the relative 2D distance between the electron and the hole that constitute the (neutral) exciton, and $K_{0}(x)=H_{0}(x)-Y_{0}(x)$, in terms of the Struve and Bessel functions of the first kind, $ H_{0}(x) $ and $ Y_{0}(x)$, respectively. This potential incorporates the dielectric environment \cite{tran,pasqual,aida} via the average dielectric permittivity $\epsilon$ (due to the substrates) in two quantities: first, it occurs in the expression of the effective Rydberg energy 
$ R_{y}=e^{2}/2\epsilon a_{B}$, in terms of the Bohr radius $a_B=\hbar^2\epsilon/e^2 \mu$ and the reduced mass $\mu=m_em_h/(m_e+m_h)$, and second it enters in the screening length $\ell_s$ [see Eqs. (\ref{eq:chi}) and (\ref{eq:scrl})]. If we neglect for the moment corrections due to the Berry curvature that are discussed in the following section, the interlayer exciton Hamiltonian therefore consists of six terms \cite{Houssem2023},

\begin{widetext}
  \begin{equation} \label{eq:eq1}
H_{ix}=\frac{p^{2}}{2\mu}+V_{RK}(r)+\frac{m_{e}^{3}+m_{h}^{3}}{2M^{2}}\omega_{cm}^{2}r^{2}+\frac{P^{2}}{2M}+\frac{M\omega_{cm}^{2}}{2}R^{2}+|m_{h}-m_{e}|\omega_{cm}^{2}\textbf{R}\cdot\textbf{r},
\end{equation}  
\end{widetext}

where
\begin{eqnarray}\nonumber
    \br=\br_e - \br_h \qquad \text{and} \qquad \frac{\bp}{\mu}=\frac{\bp_e}{m_e}-\frac{\bp_h}{m_h},\\
    \bR=\sum_{\alpha=e,h}\frac{m_{\alpha}\textbf{r}_{\alpha}}{M} \qquad \text{and}\qquad \bP=\sum_{\alpha=e,h}\textbf{p}_{\alpha}.
\end{eqnarray}
are relative and center-of-mass coordinates, respectively.

In the absence of a moir\'e potential, the Hamiltonian would only consist of the first two terms that describe the relative motion of the electron-hole compound, to which we will add the corrective terms due to the Berry curvature in the next section. Furthermore, the fourth term represents the motion of the center of mass of the exciton. The plane-wave character is modified mainly by the fifth term, which has a tendency to localize the exciton in the minima of the moir\'e potential that are approximated by parabolic potentials (harmonic approximation). The latter approximation is clearly justified as long as the moir\'e lattice spacing $L_m$ is much larger than all other length scales, namely the effective size of the exciton that is given by the Bohr radius $a_B$ that is on the order of some nm.

Within the above decomposition, we have assumed that both the electron and the hole exhibit the same potential size, leading to $m_e \omega_e = m_h \omega_h$. Furthermore, we employ the approximation $\sum_{\alpha=e,h} m_{\alpha}\omega_{\alpha}^2=M\omega_{cm}^2$, which characterizes the center-of-mass localization in terms of energies. This yields $\sum_{\alpha=e,h} \hbar\omega_{\alpha}/R_{\alpha}^{2}=\hbar\omega_{cm}/R_{cm}^{2}$. Here, we define $\omega_{cm}$ and $R_{cm}=\sqrt{\hbar/M\omega_{cm}}$ as the interlayer exciton's center-of-mass frequency and localization radius, respectively. These components can be expressed in terms of the moiré period as $\omega_{cm}=4\pi\sqrt{2V_0/M}/L_M$ and $R_{cm}=\sqrt{\ell_{cm}L_m}$, with a characteristic length $\ell_{cm}=\hbar/4\pi\sqrt{2M V_0}\simeq 1.46$\AA.

Notice finally that the moir\'e potential has also a small effect on the relative exciton motion, as it may be seen from the third term, which we explicitly take into account for a quantitative calculation of the exciton spectra for twisted bilayer samples. The last term indicates that relative and center-of-mass degrees of freedom are eventually coupled via the moir\'e potential. However, we anticipate that this term can be neglected in the case of a weak twist angle due to its proportionality to $\hbar\omega_{cm}(a_{B}/L_m)$.

In the remainder of this paper, we are interested mainly in the spectrum of the interlayer excitons, taking into account the first two terms of Hamiltonian (\ref{eq:eq1}) plus the corrections to the relative motion due to the moir\'e potential (third) term, while the effect of the center-of-mass Hamiltonian is mainly to trap the excitons in the potential minima, and the last term (coupling between the relative and center-of-mass degrees of freedom) is neglected. Most saliently, this separation into relative and center-of-mass dynamics, which cannot be achieved within the more appropriate description of the exciton in terms of a \textit{Dirac exciton} \cite{trushin}, remains possible if Berry-curvature corrections are taken into account. The latter take into account approximately the coupling between the electron and hole branches, which is natural in a description in terms of the Dirac equation and that is recalled in the following subsection.

\begin{table*}
\caption{ The input parameters used in this work, such as the lattice constant, electron and hole effective masses (in units of the free electron mass $m_{0}$), screening length, and dielectric constant, are taken from reference \cite{david}. The $2\Delta$ bandgap of $WSe_{2}$, $MoS_{2}$ monolayers (MLs), and $MoSe_{2}/WSe_{2}$ heterostructures are taken from references \cite{gua} and \cite{tran}, respectively. The table also presents the first intralayer and interlayer exciton binding energies, $E^{\tilde{n},\tilde{\ell}}_{b}$, in meV units, taking into acount  the Berry correction term.}\label{Tab:tab1}
\begin{tabular}{|c|ccccccccccc|}
\hline
\\Materials&$2\Delta[eV]$&$a_{0}$[$\AA$]&$m_{e}/m_{0}$&$m_{h}/m_{0}$&$\mu/m_{0}$&$\ell_{*}[\AA]$&$ \epsilon $&$E_{b}^{1s}$&$ E_{b}^{2p_{+}}/E_{b}^{2p_{-}} $&$E_{b}^{2s}$&$ E_{b}^{3d_{+}}/E_{b}^{ 3d_{-}}$ \\
\hline
\\ML-$WSe_{2}$&1.82&3.299 & 0.50& 0.45&0.236& 45.11&1.5&338 &192$/$204&146&117/124\\
&& & & & &&4&125 &70$/$55&37&32$/$25\\
\hline
\\ML-$MoSe_{2}$&1.72&3.286 &0.80 &0.5 &0.307&39.79&1.5&403&230$/$250 &178&142/153 \\
& & & & &&&4&155 &88$/$66&48&30/42\\
\hline
\\$ MoSe_{2}/WSe_{2} $&1.48& & &&0.288& &1.5&212 &149$/$143&111 &100/104 \\
&& & & & &&4&98 &56$/$51&37&28$/$32\\
\hline
\end{tabular}
\end{table*}

\subsection{Berry curvature of the interlayer exciton}

As discussed in the preceding subsection, our focus shifts towards the role of Berry curvature in the IX relative motion part. The non-commutativity of the position operator, stemming from the presence of non-zero Berry curvature, prompts us to adopt a generalization of the Peierls substitution \cite{Gosselin2006, Chang2008,Gosselin2008}. 
This substitution, defined by canonical coordinates, is expressed as follows: $\hat{\textbf{p}}_{\alpha}=\textbf{p}_{\alpha}$ and $\hat{\textbf{r}}_{\alpha}=\textbf{r}_{\alpha}+\boldsymbol{\Omega}_{\alpha}\times \textbf{p}_{\alpha}/2\hbar$. 
This change in coordinates results in non-commutative relations satisfying: $[\hat{\textbf{x}}_{\alpha},\hat{\textbf{y}}_{\alpha}]=i\boldsymbol{\Omega}_{\alpha}$. Consequently, new relative canonical variables are obtained: $\hat{\textbf{p}}=\textbf{p}$ and $\hat{\textbf{r}}=\textbf{r}+\left[\boldsymbol{\Omega}_{e}(\textbf{p}_{e})\times \textbf{p}_{e}-\boldsymbol{\Omega}_{h}(\textbf{p}_{h})\times \textbf{p}_{h}\right]/2\hbar$.
To assess the impact of Berry curvature on the IX relative motion Hamiltonian which includes the moiré term effect, denoted as
\begin{equation}
  \hat{H}^{ix}_{RM}(\hat{\textbf{r}},\hat{\textbf{p}})=\frac{\hat{\textbf{p}}^2}{2\mu}+V_{RK}(|\hat{\textbf{r}}|)  +\frac{m_{e}^{3}+m_{h}^{3}}{2M^{2}}\omega_{cm}^{2}\hat{\textbf{r}}^{2},
\end{equation}
 we employ a second-order Taylor expansion of the charge carriers' Berry curvature, aided by the Foldy-Wouthuysen transformation \cite{foldy}. This yields the following modified IX relative motion Hamiltonian:
\begin{equation}\label{eq:eq3}
\hat{H}^{ix}_{RM}=\dfrac{p^{2}}{2\mu}+V_{RK}(\textbf{r})+ \frac{m_{e}^{3}+m_{h}^{3}}{2M^{2}}\omega_{cm}^{2}r^{2}+H^{ix}_{B}+H^{ix}_{D},
\end{equation}
where the two corrective terms are given by

\begin{equation}\label{eq:eq4}
H^{ix}_{B}=\frac{1}{2\hbar}\boldsymbol{\nabla}_{r}V_q(r)\left[\boldsymbol{ \Omega}_{ix}(\textbf{p},\textbf{P})\times \textbf{p}+\boldsymbol{\beta}_{+} (\textbf{p},\textbf{P})\times \textbf{P}\right],
\end{equation}
and
\begin{equation}\label{eq:eq5}
H^{ix}_{D}=\frac{|\Omega_{ix}|}{4}\boldsymbol{\nabla}^{2}_{r}V_{q}(r).
\end{equation}
Here, the potential $V_q(r)=V_{RK}(r)+\frac{m_{e}^{3}+m_{h}^{3}}{2M^{2}}\omega_{cm}^{2}r^{2}$ takes into account both the effect of the direct electron-hole interaction given in terms of the Rytova-Keldysh potential (\ref{eq:eq2}) and the correction due to the moir\'e potential. We anticipate, here, that the second term is negligible with respect to the first one. Indeed, its relative weight may be estimated as 
\begin{equation}
\frac{\hbar\omega_{cm}}{R_y}    \sim \sqrt{\frac{V_0}{R_y}}\frac{a_B}{L_M},
\end{equation}
which is $\sim 0.03$ for a Bohr radius in the nm range, as compared to a characteristic moir\'e spacing of $L_m\sim 10$ nm, and where we have used $V_0\sim 10$ meV and $R_y\sim 100$ meV. The Berry cuvature's quantities  $\boldsymbol{\Omega}_{ix} (\textbf{p},\textbf{P})$ and $\boldsymbol{\beta}_{+}(\textbf{p},\textbf{P})$ are given by: 

\begin{eqnarray}\nonumber
  \boldsymbol{\Omega}_{ix} (\textbf{p},\textbf{P}) &=& \boldsymbol{\Omega}_{e}(\textbf{p}+\frac{m_{e}}{M}\textbf{P})-\boldsymbol{\Omega}_{h}(\textbf{p}-\frac{m_{h}}{M}\textbf{P}) \quad \text{and} \quad\\
  \nonumber
    \boldsymbol{\beta}_{+}(\textbf{p},\textbf{P}) &=& \frac{m_{e}}{M}\boldsymbol{\Omega}_{e}(\textbf{p}+\frac{m_{e}}{M}\textbf{P})+\frac{m_{h}}{M}\boldsymbol{\Omega}_{h}(\textbf{p}-\frac{m_{h}}{M}\textbf{P}).\\
\end{eqnarray}

The quantity $\Omega_{ix} $ appearing in Eq. (\ref{eq:eq4}) and Eq. (\ref{eq:eq5}) represents the interlayer exciton Berry curvature in the two-band model case and can be calculated analytically within the massive-Dirac-fermion model. Its derivation is presented in detail in Appendix A. The component $ \boldsymbol{\beta}_{+}(\textbf{p},\textbf{P}) $ in Eq. (\ref{eq:eq4}) is defined as the sum of two distinct Berry curvature vectors with different signs corresponding to the electron and hole contributions. We anticipate that this term will have a weaker effect compared to the term $\boldsymbol{\Omega}_{ix}(\textbf{p},\textbf{P})$ in Eq. (\ref{eq:eq4}), due to its smaller amplitude $\beta_{+}(0,0) \sim 0.2$ \AA$^2$ compared to $\Omega_{ix}(0,0) \sim 10$ \AA$^2$. Therefore, in the remainder of this paper, we will only consider the term $\boldsymbol{\Omega}_{ix}(\textbf{p},\textbf{P})$ in Eq. (\ref{eq:eq4}).

The term $H^{ix}_{B}$ appearing in Eq. (\ref{eq:eq4}), labeled as the Berry correction, can be understood as an effective spin-orbit coupling. This component can lift the degeneracy of states with angular momentum different from zero. It plays the role of an effective Zeeman effect in reciprocal space.  It is important to point out that the interlayer excitons in the present heterobilayers are formed in the vicinity of $K_{m}$ and $K_{m}^{'}$ valleys, \textit{i.e.} at points in reciprocal space where the respective Berry curvature reaches its maximal absolute values, whence the importance of the two corrective terms (\ref{eq:eq4}) and (\ref{eq:eq5}). The Berry correction term affects exciton spectra, as it couples to the gradient of Keldysh potential which is generated by the attractive interaction between the electron of $MoSe_{2}$ and the hole of $WSe_{2}$ forming the interlayer exciton states. This is a result of the Dirac character of the low-energy charge carriers in these materials, which are typically described by a massive 2D Dirac equation \cite{aida,dxiao,piechon,simon}.

In addition to the Berry correction, there is another term known as the Darwin term Eq. (\ref{eq:eq5}), where the Berry curvature couples with the Laplacian of the Rytova-Keldysh potential. This occurs in the relativistic treatment of the Dirac Hamiltonian of the two adjacent bands within the heterobilayer.

As shown in appendix \ref{appendix.a},  the charge carriers' Berry curvature  present two components: a zero order term, and a  second order component  in terms of $ \textbf{p}^{2}$ and $ \textbf{P}^{2}$, which can be neglected  [see Eq. (\ref{eq:A3}) and Eq. (\ref{eq:A4})].  The resulting interlayer exciton Berry curvature reads
\begin{equation}\label{eq:eq6}
\boldsymbol{\Omega}_{ix}\simeq\boldsymbol{\Omega}^{(0)}_{ix}(0,0)=\dfrac{\hbar^{2}}{4}\left(\dfrac{1}{m_{e}\triangle_{Mo}}+\dfrac{1}{m_{h}\triangle_{W}}\right)\vec{e}_{z}.
\end{equation}
 To understand the impact of Berry curvature on the interlayer exciton spectrum energy, we have simplified the expressions of the Berry correction and Darwin correction terms as:
\begin{eqnarray}\label{eq:eq7}
\nonumber
H^{ix}_{B} &=& \frac{i\pi R_{y}}{2 \ell^{2}_{s}}a_{B}\Omega_{ix}\frac{1}{\sqrt{r^{2}+d^{2}}}K_{1}\left(\frac{\sqrt{r^{2}+d^{2}}}{\ell_{s}}\right)\frac{\partial}{\partial\theta}\\
&&+ i\hbar\omega_{cm}\frac{\alpha_{0}}{2}\frac{\Omega_{ix}}{R_{cm}^2}\frac{\partial}{\partial\theta},
\end{eqnarray}
and
\begin{equation}\label{eq:eq8}
H^{ix}_{D} = \frac{\Omega_{ix}\pi R_{y}}{4\ell^{2}_{s}}\Xi(r)
+\hbar\omega_{cm}\frac{\alpha_0}{4}\frac{\Omega_{ix}}{R_{cm}^{2}}.
\end{equation}
Here, the quantity $\alpha_0=(m_e^3+m_h^3)/M^3$ is related to the (less relevant) contribution of the exciton moir\'e potential to the corrective terms, 
and the term $\Xi(r)$ appearing in Eq. (\ref{eq:eq8}) is given by

\begin{eqnarray}
\nonumber
\Xi(r) &=& \frac{a_{B}r^{2}}{\ell_{s}(r^{2}+d^{2})}K_{0}\left(\frac{\sqrt{r^{2}+d^{2}}}{\ell_{s}}\right)\\
&&+\frac{a_{B}(d^{2}-r^{2})}{(r^{2}+d^{2})^{\frac{3}{2}}}K_{1}\left(\frac{\sqrt{r^{2}+d^{2}}}{\ell_{s}}\right). 
\end{eqnarray}
The function $K_0(x)$ has already been defined in the context of the Rytova-Keldysh potential (\ref{eq:eq2}), and we have used $ K_{1}(x)=H_{1}(x)-Y_{1}(x)$, where $H_{1}(x)$ and $Y_{1}(x)$ are the first-order Struve and Bessel functions of the second kind, respectively.

\subsection{Diagonalization of the exciton Hamiltonian and exciton wave functions}

The diagonalization of the Hamiltonian matrix, constructed from Eq. (\ref{eq:eq3}) with the Rytova-Keldysh potential Eq. (\ref{eq:eq2}) and with the corrective terms Eq. (\ref{eq:eq7}) and Eq. (\ref{eq:eq8}), in the 2D-hydrogenic basis state  $|\varphi_{n,\ell}\rangle$,  leads to the following eigenvalues $E_{\tilde{n},\tilde{\ell}}$ and their corresponding eigenvectors 
\begin{equation}
|\psi_{\tilde{n},\tilde{\ell}}\rangle=\sum_{n,|\ell|<n}A^{\tilde{n},\tilde{\ell}}_{n,\ell}|\varphi_{n,\ell}\rangle. 
\end{equation}
Here, $A^{\tilde{n},\tilde{\ell}}_{n,\ell} $ are their coefficients obtained after numerical diagonalization. The corresponding basis state used in this work has the following expression
\begin{equation}
\varphi_{n,\ell}(r,\theta)=\dfrac{Y_{n,\ell}(r)e^{i\ell\theta}}{\sqrt{2\pi}}, 
\end{equation}
where the radial part 
\begin{equation}
Y_{n,\ell}(r)=\frac{C_{n,\ell}}{a_B}e^{-\frac{r\alpha_{n}}{2a_{B}}}\left(\alpha_{n}\frac{r}{a_{B}}\right)^{|\ell|}L_{n-|\ell|-1}^{2|\ell|}\left(\alpha_{n}\frac{r}{a_{B}}\right),
\end{equation}
is given in terms of associated Laguerre polynomials $ L^{a}_{n}(x)$, and the normalization constants $C_{n,\ell}$ read
\begin{equation}
C_{n,\ell}=\frac{4}{(2n-1)^{\frac{3}{2}}}\left(\frac{(n-|\ell|-1)!}{n+|\ell|-1)!}\right)^{\frac{1}{2}}.
\end{equation}
The terms $n$ and $\ell$  are  the principal quantum number and the angular momentum, respectively, with $ n=1,2,3... $ and $-(n-1)\leq\ell\leq n-1 $. The states are $(2n-1)$-fold degenerate, and we recall that they are currently labeled \textit{s} for
$ \ell=0 $, \textit{p} for $ \ell=\pm 1 $ and \textit{d} for $ \ell=\pm 2 $. The index numbers $\tilde{n}   $ and $ \tilde{\ell} $ labeling the eigenstates refer as well to the dominant contribution of the coefficients $ A^{\tilde{n},\tilde\ell}_{\bar{n},\bar{\ell}} $.

To illustrate the present framework, we have identified several key figures that capture the main results. These will be presented in the following section.

\section{Results and discussions}\label{sec3}

We now present the results for the spectra obtained by numerical diagonalization of the exciton Hamiltonian. We are mainly interested in the effect of the Berry curvature on the binding energies of the interlayer excitons (IX), denoted as $E_{b}^{\tilde{n},\tilde{\ell}}=-E_{\tilde{n},\tilde{\ell}}$. The different splittings of the exciton lines are investigated as a function of the interlayer distance and dielectric environment, for zero twist angle between the layers. The effect of the twist on the exciton spectra will be discussed separately, as well as the role of the photon polarization. 

\subsection{IX binding energies spectra: effects of dielectric environment and interlayer separation on  $|E_{1s}-E_{2p_{\pm}}|$ and $|\Delta E_{2p_{\pm}}|$ splitting energies}

\begin{figure}
\begin{center}
 \includegraphics[scale=0.35]{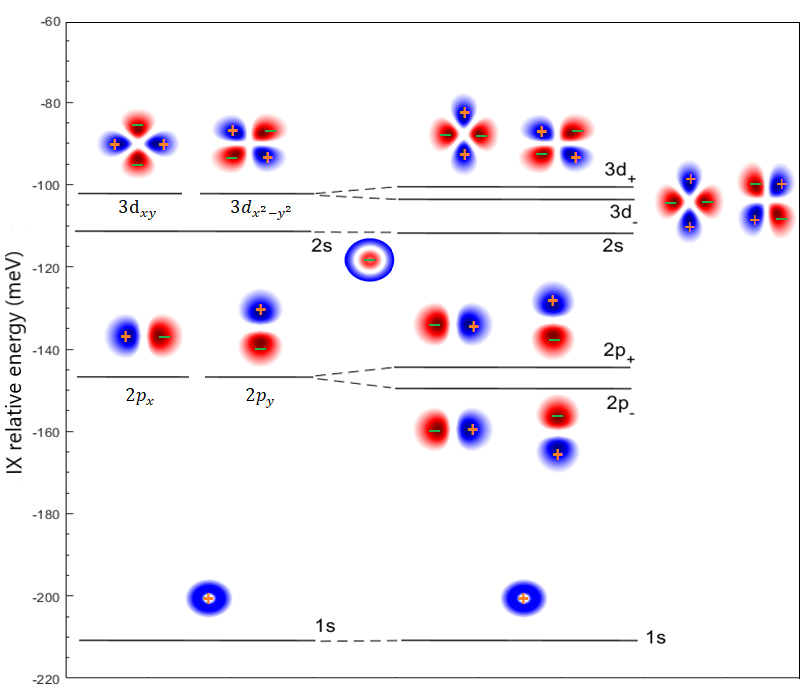}
 \\
 \includegraphics[scale=0.21]{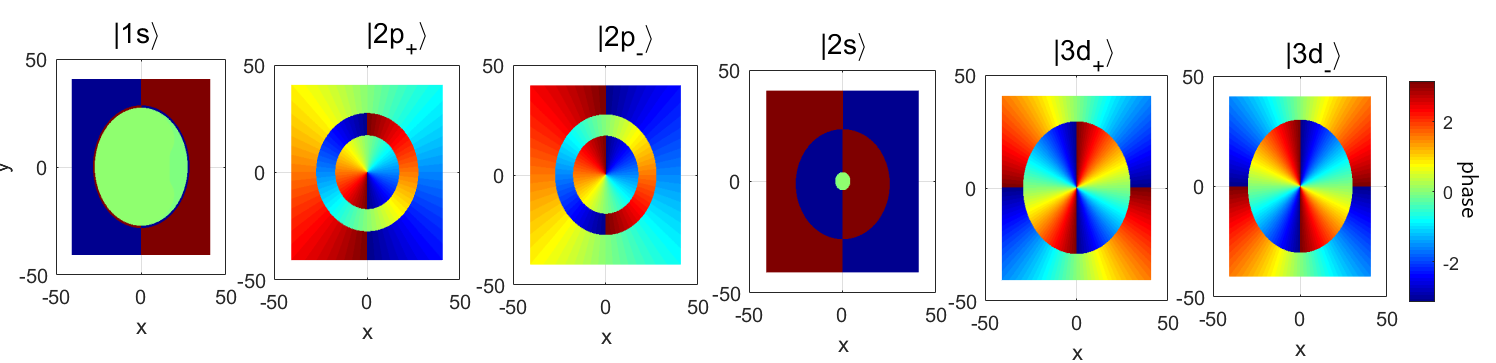}
 \caption{ The  interlayer exciton eigenvalues spectra,  $E_{\tilde{n},\tilde{\ell}}$,  within the $R-\text{MoSe}_{2}/\text{WSe}_{2}$ heterobilayer and deposited on a $\text{SiO}_{2}$ substrate ($\epsilon_{\text{SiO}_{2}} = 2.1$) is presented in units of meV, considering an interlayer distance of $d = 7 \text{\AA}$, with a negligible twist between the layers. On the left hand side, the energies of the Rydberg states \textit{1s}, \textit{2p}, \textit{2s} and \textit{3d}, along with their corresponding eigenvectors, are displayed without taking into account Berry-curvature corrective terms, $\boldsymbol{\Omega}_{ix}=0$. The effect of the Berry curvature on the exciton spectra, via the corrective terms (\ref{eq:eq7}) and (\ref{eq:eq8}), is shown on the right hand side. Mainly the pseudospin-orbit-type term (\ref{eq:eq7}) yields a degeneracy lifting of the $2p_{\pm}$ and $3d_{\pm}$ dark states and subsequent modifications in their respective eigenvectors. Here, the color scheme distinguishes positive values in blue and negative values in red. The Darwin-type term (\ref{eq:eq8}) yields a slight shift of the energies of the $ns$ states, in the meV range that is barely visible in the presented energy range. Below the spectrum, the corresponding complex phase of the first six eigenstates is visualized, revealing a zero phase for $|ns\rangle$ states and highlighting the twofold rotational degree of freedom exhibited by $|2p_{\pm}\rangle$ and $|3d_{\pm}\rangle$ states.
}\label{fig2}
 \end{center}
\end{figure}

\begin{figure*}
\begin{center}
 \includegraphics[scale=0.63]{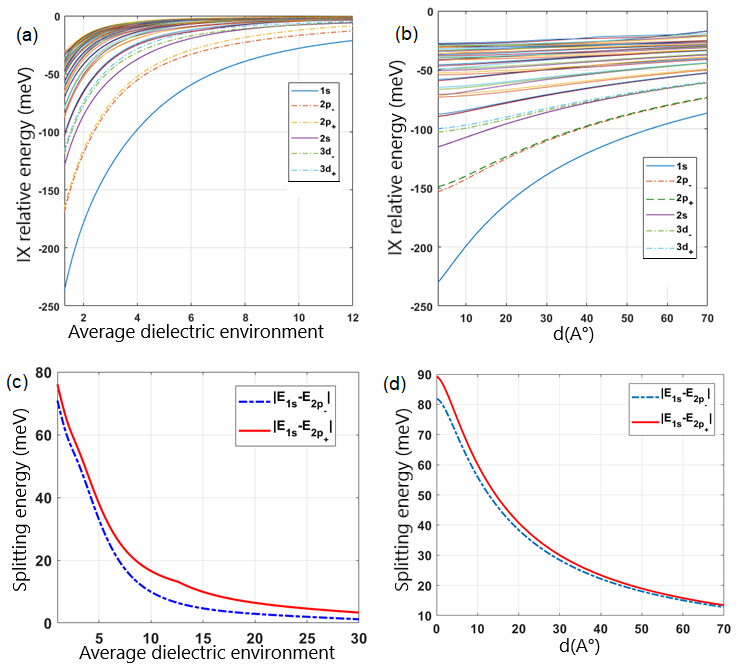}
 \caption{Dependence of the interlayer exciton relative energies spectra on the average dielectric environment $ \epsilon $ [for a fixed distance $d=7$ \AA, panel (a)] and on the interlayer distance $ d $ [for a fixed value of $\epsilon=1.5$, panel (b)], respectively, displaying 144 eigenvalues. Panels (c) and (d) depict the dependence of the energy difference $ |E_{1s}-E_{2p_{\pm}}| $ for the two circular polarizations $\pm$ on  $ \epsilon $ and $ d $, respectively.}\label{Fig3}\end{center}
\end{figure*} 

\begin{figure*}
\begin{center}
 \includegraphics[scale=0.33]{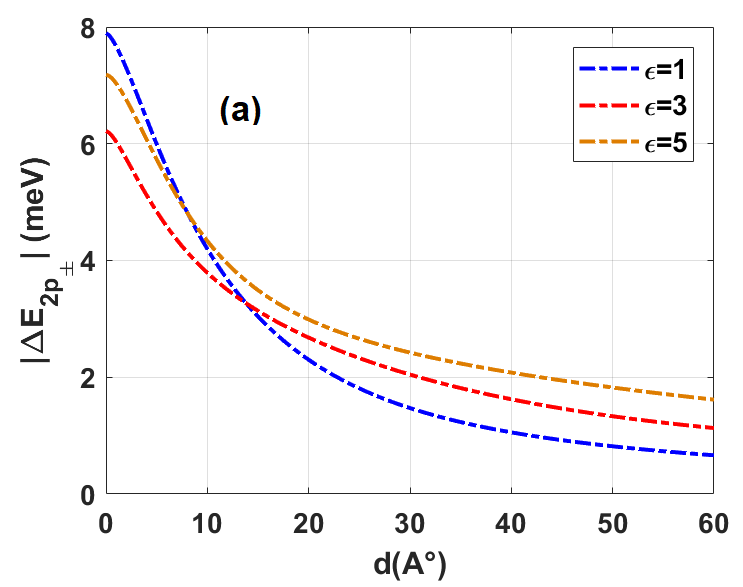}
 \quad
 \includegraphics[scale=0.3]{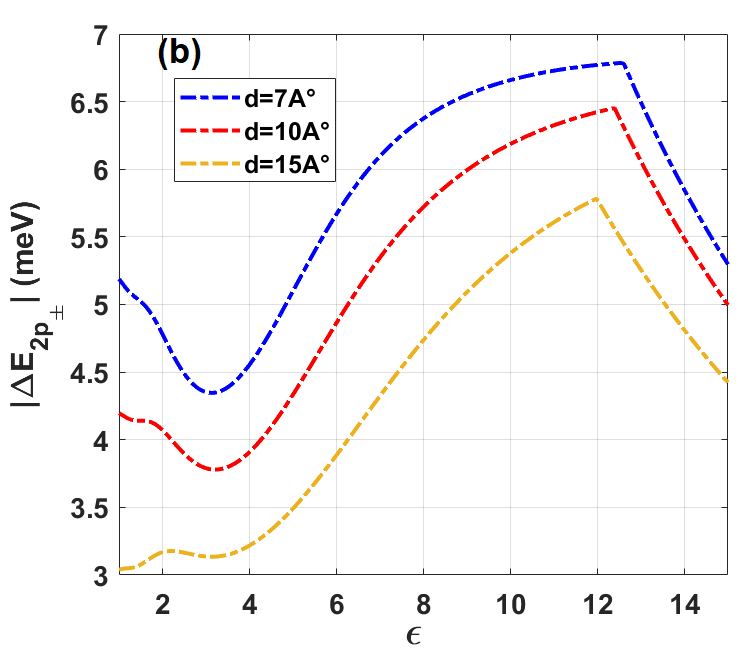}
 \caption{Dependence of the splitting $ |\Delta E_{2p_{\pm}}| $ between the two circular polarizations of the $2p$ states on the layer separation [for a fixed value of $\epsilon=1.5$, panel (a]) and the average dielectric environment [for a fixed layer separation of $d=7$ \AA, panel (b)], respectively.}\label{Fig4}\end{center}
\end{figure*}

\begin{figure*}
\begin{center}
 \includegraphics[scale=0.41]{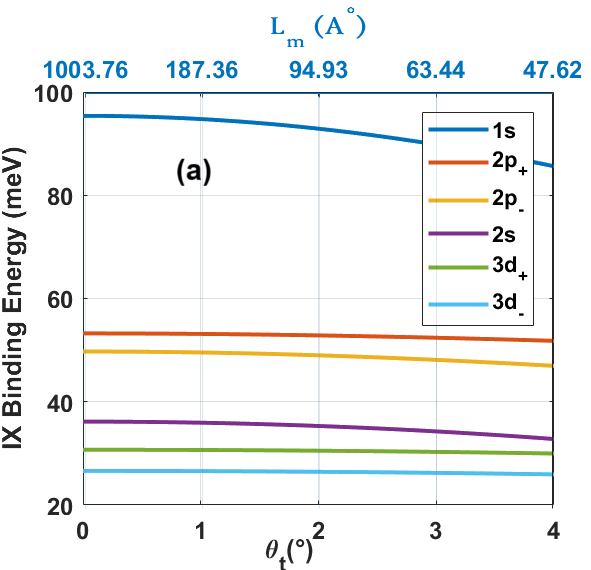}
 \quad
  \includegraphics[scale=0.4]{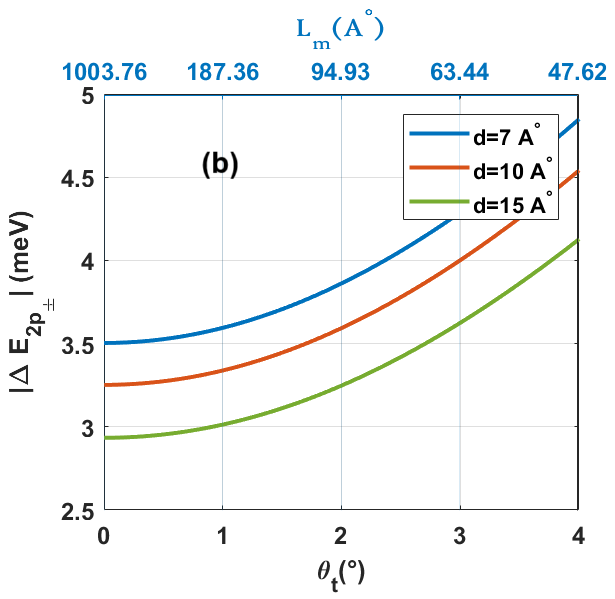}
 \caption{Figure (a) depicts the twist angle dependence of IX binding energies within $MoSe_2/WSe_2$ heterobilayer encapsulated by hBN. Figure (b) shows the twist angle dependence of $ |\Delta E_{2p_{\pm}}| $ splitting energy, taken for different value points of the spacing separation d.}\label{Fig.lm}\end{center}
\end{figure*}

  

Figure \ref{fig2} shows the energy levels of the interlayer exciton Rydberg states \textit{1s}, \textit{2p}, \textit{2s}, and \textit{3d}, along with their corresponding eigenvectors $|\psi_{\tilde{n},\tilde{\ell}}\rangle$, within the $R-\text{MoSe}_{2}/\text{WSe}_{2}/\text{SiO}_2$ heterostructure. In the calculation of the spectra, we have considered perfect alignment between the layers so that the effect of the twist is negligible ($L_m\rightarrow \infty$ and thus $\omega_{cm}\rightarrow 0$). Furthermore, we have used physically relevant values for the interlayer distance $d=7$ \AA and the dielectric environment $\epsilon=1.5$ corresponding to a SiO$_2$ substrate with a dielectric constant of  $\epsilon_{\text{SiO}_{2}} = 2.1$.
The left column of Fig. \ref{fig2} shows the energy levels without incorporating the Berry curvature terms, while the energies on the right hand side take into account the latter via the terms (\ref{eq:eq7}) and (\ref{eq:eq8}). Most saliently, we observe a splitting of the degenerate Rydberg states, such as the $2p_{\pm}$ and $3d_{\pm}$ dark states. We observe a splitting energy of $|\Delta E_{2p_{\pm}}|=|E_{2p_{+}}-E_{2p_{-}}|=5$ meV for interlayer exciton states, akin to the behavior of a spontaneous Zeeman effect, attributable to the Berry correction term (\ref{eq:eq7}). Notice that the Darwin term (\ref{eq:eq8}) has a much weaker effect on the exciton spectra. Indeed, it is extremely local -- in the case of a pure Coulomb potential $\sim 1/r$, Eq. (\ref{eq:eq5}) yields a Dirac delta term, $\nabla_r^2 V(r)\propto \delta(r)$ -- and therefore mainly affects the $ns$ states, which have a non-zero amplitude at the origin. However, even in the case of the most prominent $1s$ state, its energy is only shifted by $1.5$ meV for the parameters used in our calculations, \textit{i.e.} a shift of only $\sim 0.7\%$ as compared to its binding energy.

The relevant parameters used in our calculations are given in Table \ref{Tab:tab1}. Here, we also show the binding energies of the intralayer excitons for WeSe$_2$ and MoSe$_2$ monolayers, for comparison. The energy splitting between their $2p_+$ and $2p_-$ states is approximately 12 meV and 20 meV, respectively. The observed splitting energy for interlayer excitons is therefore lower compared to that for intralayer excitons, possibly due to the smaller spacing between the layers, leading to the excited states of interlayer excitons being closely situated, thereby creating a denser spectrum. These findings agree well with previously reported results \cite{trushin,Chaw,aida,Tang2023,Wu2015}.

Figures \ref{Fig3}(a) and \ref{Fig3}(b) illustrate the dependence of interlayer exciton energies on the average dielectric environment and spacing separation, respectively, displaying 144 eigenvalues. In panel (a), where we vary the dielectric constant $\epsilon$, we have used a fixed value of $d=7$ \AA, while the variation of the interlayer distance $d$ in panel (b) is carried out for a fixed value of $\epsilon=1.5$.

They clearly demonstrate a significant decrease in binding energies with an increase in either $ \epsilon $ or $d$, as it is expected from the dependence of the effective potential on these parameters. Notice that the original $1/r$ potential is replaced by $1/\sqrt{r^2+d^2}$ due to the layer separation and thus the separation between the electron in one layer and the hole in the other one. Furthermore, the binding energies are controlled by the Rytova-Keldysh potential (\ref{eq:eq2}), which scales as $\sim \ell_s/r \sim 1/\epsilon$ in the limit $\ell_s\rightarrow 0$  [\textit{i.e.} $\epsilon\rightarrow \infty$, see Eqs. (\ref{eq:chi}) and (\ref{eq:scrl})].
The dependence of IX splitting energies $ |E_{1s}-E_{2p_{\pm}}| $ on $ \epsilon $ and $ d $ is also depicted in figures \ref{Fig3}(c) and \ref{Fig3}(d), respectively. These figures reveal a considerable decrease in spacing energy, tending toward the range of Terahertz frequencies as $ \epsilon $ or the spacing separation between the two layers increases. These findings may have significant implications for the development of high-performance ultrafast devices.

 Figures \ref{Fig4}(a) and \ref{Fig4}(b) represent the dependency of $ |\Delta E_{2p_{\pm}}| $, the interlayer exciton splitting energy, on spacing separation and average dielectric environment for various values of $ \epsilon $ and $ d $, respectively. Figure \ref{Fig4}(a) shows a decrease in spacing energy with increasing $d$, as one may have anticipated from the scaling arguments mentioned above. Indeed, the splitting energy increases when both $d$ and $ \epsilon $ approach weak values. As shown in Fig. \ref{Fig4}(a), $ |\Delta E_{2p_{\pm}}| $ reaches a maximum value of approximately $ 8$ meV when $ \epsilon=1 $ and $d$ is close to zero, consistent with findings reported by Tang et al. \cite{Tang2023}.

 It is noteworthy that adjusting the spacing $d$ between the two monolayers of TMDs can be achieved by incorporating layers of hBN, a typical experimental technique. This interplay can significantly impact the interlayer band-gap, although it does not notably affect the effective masses. Conversely, analyzing $ |\Delta E_{2p_{\pm}}| $ as a function of the average dielectric environment, taken from different values of interlayer distance, reveals an unexpected non-monotonic behavior as depicted in Fig. \ref{Fig4}(b). We identify three distinct regimes for the splitting $ |\Delta E_{2p_{\pm}}| $: a decreasing trend within $\epsilon$ $ \in $ [1, 3] when $d=7$ \AA, followed by an increasing trend within $ \epsilon $ $ \in $ [3, 12.8], and a subsequent decrease within $ \epsilon $ $ \in $ $[12.8, \infty]$. The latter is expected again based on the scaling arguments. 
 
 While we do not provided a full-fletched theoretical explanation of this non-monotonic behavior on the dielectric function, we would emphasize that it is likely due to the hidden dependence of the effective interaction potential on the dielectric constant, which enters as a global scaling factor only in the large-$\ell_s$ limit, as mentioned above. In an intermediate regime, one needs to take into account that the Rytova-Keldysh potential is itself a (non-monotonic) function of the screening length $\ell_s\sim \epsilon$. Notice finally that this non-monotonic behavior cannot be obtained when using pure Coulomb interaction, which yields $|\Delta E_{2p_{\pm}}|=\frac{64}{81} \frac{\Omega_{ix}}{a_{B}^{2}}R_{y}$, as reported in \cite{aida, zhou}. Consequently, this suggests a decrease in splitting energy as the dielectric environment increases, owing to its proportionality with $ 1/\epsilon^{4} $.



\subsection{Role of the moir\'e potential in the IX energies}
 
 In the previous section, we have not taken into account the role of the twist between the two layers on the exciton energies, \textit{i.e.} we have considered a situation with a twist angle near zero degree. Within this case, the quadratic potential which incorporates the moiré traps effects has not a significant effect on IX energy spectra, where it shifts the ground state by approximately 1 meV. Thus, to assess the impact of moiré  effect on IX energies, Fig. \ref{Fig.lm}(a) shows the dependence of the first interlayer exciton Rydberg states on the twisting angle effects, which reveals a decreasing in IX binding energies as increasing the twist angle from 0° to 4° with more significant impact on the s-type Rydberg states.

 In counterpart, Fig. \ref{Fig.lm}(b) showcases 
 the dependence of twisting effect on the $|\Delta E_{2p_{\pm}}|$ splitting energy. This result reveals a considerably increasing in the splitting energy between $ 2p_+ $ and $ 2p_-  $ Rydberg states as increasing the twist angle from 0° to 4°.

 The Berry curvature does not only modify the exciton binding energy spectrum but also the corresponding eigenvectors ($\psi_{2p_{\pm}}$ and $\psi_{3d_{\pm}}$) by mixing the $2p_{x}$ and $2p_{y}$ degenerate states in the linear combination $|\psi_{2p_{\pm}}\rangle=\frac{1}{\sqrt{2}}[|\psi_{2p_{x}\rangle}\pm i |\psi_{2p_{y}\rangle}]$. The exciton $nd_{\pm}$ states are also known to exhibit dark properties in linear optics, necessitating visualization through methods such as Third Harmonic Generation (THG) \cite{Peres2021}. The $|\psi_{3d_{\pm}}\rangle$ eigenstate will be also superposed on $| \psi_{3d_{xy}\rangle}$ and $|\psi_{3d_{x^2-y^2}}\rangle$ states. The $2p_{+}$ and $2p_{-}$ exciton states have the same real part sign as $2p_{x}$ and opposite imaginary part signs of $2p_{y}$ (refer to Fig. \ref{fig2}). As a consequence, the two new eigenstates will exhibit opposing rotations due to symmetry, presenting an opportunity to explore the $1s-2p_{\pm}$ transition through the utilization of circularly polarized optical absorption originating from the $1s$ ground state. This will be the focal point of our discussion in the subsequent subsection.

\subsection{$ 1s-2p_{\pm} $ IX transition polarizability and its quantum-mechanical interaction with photons field.}
\begin{figure*}
\begin{center}
 \includegraphics[scale=0.65]{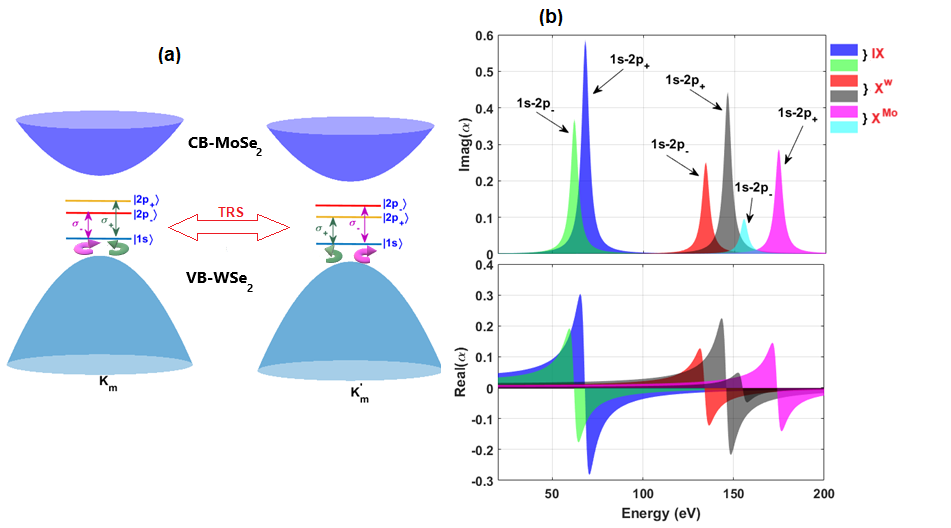}
 \caption{Figure(a) illustrates the band alignment schematics for a $ WSe_{2}/MoSe_{2}/SiO_{2} $  heterobilayer around the $K_{m}$ and $K_{m}^{'}$ valley points at zero twist angle, showcasing the transition scheme of the $1s$-$2p_{\pm}$ interlayer exciton and demonstrating the effect of time reversal symmetry (TRS).
 Figure(b) depicts the $1s$-$2p_{\pm}$ transitions' polarizability within the $WSe_{2}/MoSe_{2}/SiO_{2}$ heterobilayer for IX and intralayer exciton within $MoSe_2(WSe_2)$ monolayer, as denoted $X^{Mo(W)}$. The upper and lower panels correspond to the imaginary and real parts of the polarizability, respectively.}\label{Fig5}
\end{center}
\end{figure*}

The significant contribution of the Berry curvature lies in its ability to leverage the lifting of the $2p_{\pm}$ degeneracy, offering two degrees of freedom generated by optical circular polarization. This opens up a novel avenue in quantum information and THz devices. Moreover, harnessing the $1s-2p_{\pm}$ interlayer exciton transition in TMD heterobilayers holds immense promise across various domains including valleytronics,  and bio-sensing. These examples merely scratch the surface of the myriad potential applications stemming from the utilization of the $1s-2p_{\pm}$ interlayer exciton transition in TMDs heterobilayers.

\begin{figure*}
\begin{center}
 \includegraphics[scale=0.65]{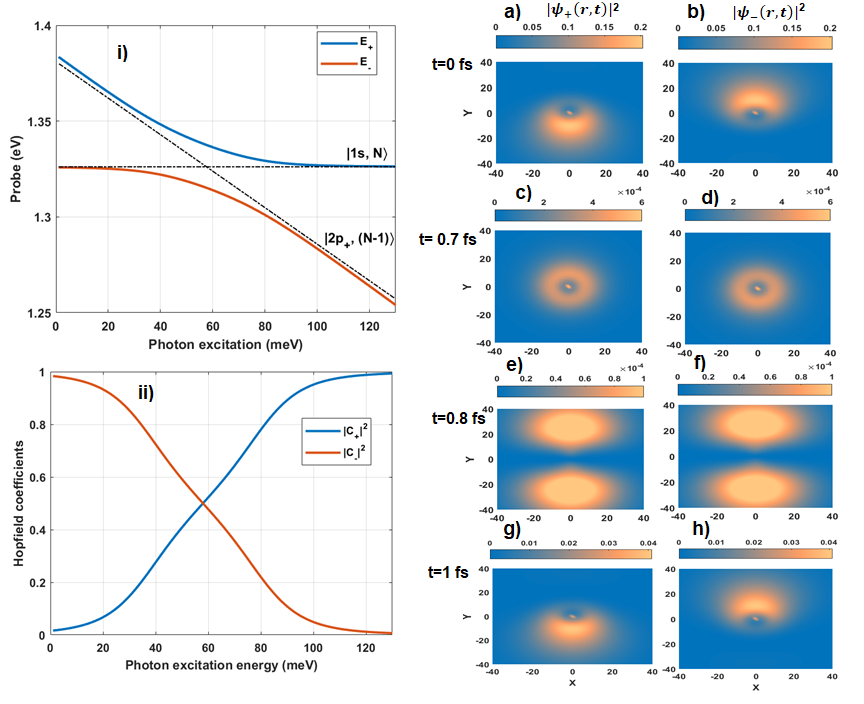}
 \caption{i) The avoiding-crossing behaviour due to quantum coupling between the  photons field and the $1s–2p_{+}$ interlayer exciton electronic transition, which is determined by using the Eq.\ref{eq:eq11}. ii) The corresponding Hopfield coefficient as a function of photon energy. (a,c,e,f) and (b,d,f,h) figures are the $|\psi_{+}(r,t)|^{2}  $ and $| \psi_{-}(r,t)|^{2} $ density, respectively, taken for different time points.}\label{Fig6}
 \end{center}
\end{figure*}

Among the well-known techniques used to observe exciton dark states in TMD monolayers, pump-probe exciton spectroscopy stands out as a highly effective method in nonlinear optics, particularly for elucidating the brightening of exciton $np_{\pm}$ dark states \cite{Peres2021, merkl}. By pumping the exciton ground state with a near-infrared (NIR) frequency, the ground state can then be probed by a medium infrared (MIR) frequency. This enables the $1s-2p_{\pm}$ exciton transition which falls within the TeraHertz frequency range for interlayer excitons in TMD heterobilayers \cite{merkl}. In this study, we aim to uncover the exciton transition between the state $1s-2p_{\pm}$ for both interlayer and intralayer excitons. We consider a circularly polarized electric field \textbf{F} in interaction with interlayer exciton ground state. We analyze the interaction Hamiltonian $H_{int}=e\hat{\textbf{r}}\cdot\textbf{F}$, where the exciton dipole moment operator is $ \hat{\textbf{r}}=\textbf{r}+\boldsymbol{\Omega_{ix}}\times \textbf{p} /2\hbar$. The electromagnetic-field interaction can be re-expressed as $\hat{\textbf{r}}\cdot\textbf{F}=\hat{\textbf{r}}^{+}\cdot\textbf{F}^{+}+\hat{\textbf{r}}^{-}\cdot\textbf{F}^{-}$, where $\hat{\textbf{r}}^{\pm}=(\hat{\textbf{r}}_{x}\pm i\hat{\textbf{r}}_{y})/\sqrt{2}=\textbf{r}^{\pm}\pm \boldsymbol{\Omega}_{ix}\times \textbf{p}^{\pm}/2\hbar$ and $\textbf{F}^{\pm}=(\textbf{F}_{x}\pm i\textbf{F}_{y})/\sqrt{2}$ is the circular polarized electric field. The dipole moment then becomes 
\begin{equation}
\hat{\textbf{r}}^{\pm}=\frac{e^{\pm i \theta}}{\sqrt{2}}\left[ r\pm \frac{i \Omega_{ix}}{2}\frac{1}{r}\frac{\partial}{\partial \theta}\right]. 
\end{equation}

To determine the exciton polarizability, we  calculate the dipole moment matrix element, which read

\begin{multline}\label{eq:eq9}
\langle \psi_{\tilde{n}^{'},\tilde{\ell}^{'}}|\hat{\textbf{r}}^{\pm}|\psi_{\tilde{n},\tilde{\ell}}\rangle=
\delta_{|\ell-\ell^{'}|,1} \sum_{n,\ell,n^{'},\ell^{'}} A^{\tilde{n},\tilde{\ell}}_{n,\ell}A^{\tilde{n^{'}},\tilde{\ell}}_{n^{'},\ell^{'}}
\\
\times\int_{0}^{\infty}Y_{n',\ell'}(r)\left[ \frac{r^2}{\sqrt{2}}\mp\dfrac{\ell\Omega_{ix}}{2\sqrt{2}}\right]Y_{n,\ell}(r)dr,
\end{multline}

the dipole moment matrix element Eq. (\ref{eq:eq9}) only exists between states with different symmetries (such as $s$ and $p$ states) and its magnitude increases as the difference between $\tilde{n}$ and $\tilde{n'}$ increases. Meanwhile, to understand the role of Berry curvature in the transition between the $1s$ and $2p_{\pm}$ states, it is important to note that around the $K_{m}$ valley point, the transition of the $1s$ to $2p_{\pm}$ interlayer exciton is exclusively influenced by coupling to $\sigma^{\pm}$ circularly polarized light. Conversely, the $K_{m}^{'}$ valley, connected to the $K_{m}$ valley via time reversal symmetry (TRS), results in an equivalent transition of $1s$ to $2p_{\mp}$ interlayer exciton state, coupling solely to $\sigma^{\mp}$ polarized light. 
This relationship is illustrated in the accompanying figure \ref{Fig5}(a).

The oscillator strength of the $ 1s-np_{\pm} $ transition is defined as \cite{merkl}:
\begin{equation}
f_{1s-np_{\pm}}=\dfrac{2\mu}{\hbar^{2}}|E_{1s}-E_{np_{\pm}}|\langle \psi_{1s}|\hat{\textbf{r}}^{\pm}|\psi_{np_{\pm}}\rangle |^{2}    
\end{equation}

Using the classical sum-over-states method, the frequency-dependent polarizability due to transitions from the $1s$ exciton state to $np_{\pm}$ states can be expressed as \cite{karplus}:
\begin{equation}
 \alpha(\hbar\omega)=\dfrac{e^{2}\hbar^{2}}{\mu}\sum_{n,\ell} \dfrac{f_{1s-np_{\pm}}}{|E_{1s}-E_{np_{\pm}}|^{2}-(\hbar\omega)^{2}-i \hbar\omega\gamma}   
\end{equation}

where $\gamma$ is a damping term that controls the line width of the resonances and is equal to 5 meV \cite{Peres2021}. 
The imaginary and real parts of the exciton polarizability (in atomic units) provide information about absorption and refractivity for each transition, respectively. We have determined the exciton polarizability for both interlayer and intralayer excitons, as shown in figure \ref{Fig5}(b), and found that the peak absorption of the $1s-2p_{\pm}$ interlayer excitons is more significant than in the intralayer excitons case. The resonant peak of the real and imaginary part of polarizability corresponding to the $\vert E_{\tilde{1s}} - E_{\tilde{2p}_{\pm}}\vert$ energy showed a high peak absorption for interlayer excitons compared to intralayer excitons. This can be especially explained by the short energy separation between $\tilde{1s}$ and $\tilde{2p}_{\pm}$ states.

 Following to our  previous results, we will now thoroughly investigate the interlayer exciton two-level system coupled with photons field. Let us first consider the initial state, denoted as $ |1s, N\rangle $, and the final state, denoted as $ |2p_{\pm}, N-1\rangle $. Here, the integer number $ N $ is defined as $ N = aa^{\dagger} $, where $ a $ ($ a^{\dagger} $) represents the annihilation (creation) photon operator. This system can be described by the following Hamiltonian \cite{Chaw}: 
\begin{equation}\label{eq:eq10}
H_{i}=\begin{pmatrix}
    E^{1s}_{IX} & \dfrac{V_{int}}{2} \\
    \dfrac{V_{int}}{2} & E^{1s}_{IX}- \Delta_{0} \\
\end{pmatrix} .
\end{equation}
Here $ \Delta_{0}=E_{ph}-|E_{IX}^{1s}-E^{2p_{\pm}}_{IX} |$ is the detuning energy. $ E_{IX}^{1s(2p_{\pm})}=2\Delta-E_{b}^{1s(2p_{\pm})} $ is the interlayer exciton ground state ($ 2p_{\pm} $ state) energy. $ E_{ph}=\sum\hbar\omega(N+\frac{1}{2}) $ is the photon excitation energy, describes the pump process.
 The coupling term  $ V_{int} $ is associated with
 the probe process, which is proportional to $ 1s-2p_{\pm} $ interlayer dipole moment $ \hat{\textbf{r}}_{1s-2p_{\pm}} $, througth the expression $ V_{int}= \boldsymbol{ \hat{r}}_{1s-2p_{\pm}}\cdot \textbf{F}^{\pm}$. By directly diagonalized the Hamiltonian shown in Eq. (\ref{eq:eq10}), we can obtain the
eigenenergies:

\begin{equation}\label{eq:eq11}
E_{\pm}=E_{IX}^{1s}-\frac{\Delta_{0}}{2}\pm\frac{\sqrt{V_{int}^{2}+\Delta_{0}^{2}}}{2} .
\end{equation}
  The corresonding eigenvectors are given by: 
\begin{equation}\label{eq:eq12}
|\psi_{\pm}\rangle= C_{\pm}|1s,N\rangle+C_{\mp}|2p_{\pm},N-1\rangle
\end{equation}

Here, $ C_{\pm} $ is the Hopfield coefficient satisfying $ |C_{+}|^{2}+|C_{-}|^{2}=1 $, with 
\begin{equation}
|C_{\pm}|^{2}=\frac{1}{2}\left( 1\pm \frac{\Delta_{0}}{\sqrt{\Delta_{0}^{2}+V_{int}^{2}}}\right) . 
\end{equation}
As anticipated,  Eq. (\ref{eq:eq11}) illustrates the quantum mechanical coupling between matter and photon-dressed states, manifesting as energy-level repulsion. This interaction leads to the emergence of upper and lower exciton bands. Figure \ref{Fig6}(i) delineates these two repulsive branches, while their respective Hopfield coefficients are elucidated in Figure \ref{Fig6}(ii). Notably, the unmistakable anticrossing between the bands is evident around $\Delta_{0}=0$ detuning energy, indicating an anticipated self-hybridization between the $1s$ and $2p_{+}$ states influenced by photon fields.

To assess the hybridization of the $1s$ and $2p_{+}$ orbitals under the influence of an optical Stark effect, we scrutinize the dynamics of the upper and lower exciton bands. Utilizing the time-dependent eigenvectors: $\psi_{\pm}(r,t)=e^{-i(E_{+}+N\hbar\omega)\frac{t}{\hbar}} C_{\pm}|1s,N\rangle \pm e^{-i(E_{-}+(N-1)\hbar\omega)\frac{t}{\hbar}}C_{\mp} |2p_+,(N-1)\rangle $, we present in figures \ref{Fig6}(a-h) the evolving density $|\psi_{\pm}(r,t)|^{2}$ under strong regime coupling at $\Delta_{0}=0$, observed at various time points. These figures distinctly illustrate the mixing of orbitals, showcasing ultrafast transitions occurring within the femtosecond range. The two-level dressed model employed in this study to describe the optical Stark effect holds substantial significance in optics, offering valuable insights into the interaction between the $1s-2p_{\pm}$ interlayer exciton transition and photon fields. Moreover, the Berry curvature assumes a pivotal role by providing two degrees of freedom can be generated by circular optical polarization.

\section{Conclusion}\label{sec5}

In conclusion, we have studied the influence of the Berry curvature on the energy levels of interlayer excitons in the $R$-stacked WSe$_{2}$/MoSe$_{2}$ heterobilayers. The Berry curvature arises in two terms affecting the exciton spectra. The Darwin term only shifts the $ns$ levels by a small energy in the meV range, while the term, which is reminiscent of spin-orbit coupling, has a more prominent effect in that it splits the $2p$ (and $3d$) levels. The splitting, as well as the exciton spectra, depend on several external parameters such as interlayer separation, the dielectric environment and the twist angle between the TMD layers. The obtained $2p_\pm$ splitting is in the range of $3...8$ meV, and thus a bit lower than for their monolayer (or intralayer) counterparts. Generally, the transitions $|E_{1s} - E_{2p_{\pm}}|$ are reduced as the parameters $\epsilon$ or $d$ are increased. This observation holds promise for the advancement of high-performance, high-speed devices. Furthermore, we observed a decrease in the splitting energy, $|\Delta E_{2p_{\pm}}|$, as the interlayer distance $d$ is increased, in line with our expectations. Surprisingly, when assessed as a function of $\epsilon$, this variation revealed three distinct regimes, probably as a consequence of the non-monotonic dependence of the interaction potential on the screening length, which itself depends on $\epsilon$. We have also investigated the interlayer polarizability of the $1s$-$2p_{\pm}$ transition, employing a two-level dressed model to explore the optical Stark effect. In summary, our results are in strong agreement with prior studies and hold significant potential for the design of ultrafast devices.

\section*{Acknowledgements}

H.E.H and S.J. acknowledge financial support from Tunisian Ministry of Higher Education and Scientific Research and also the financial support from the Shemera project.
H.E.H and M.O.G. acknowledge financial support from the French National Research Agency (project TWISTGRAPH) under Grant No. ANR-21-CE47-0018. 

\begin{widetext}
\appendix
\section{ Electron(hole) Berry curvature}\label{appendix.a}

In the framework of the massive Dirac fermion model, the Berry curvature of Bloch states for charge carriers exhibits the following characteristic form: \cite{zhou,piechon}:

\begin{equation}\label{eq:A1}
\boldsymbol{\Omega}_{e(h)}(\textbf{p}_{e(h)})=\pm \dfrac{\hbar^{2}\alpha_{Mo(W)}^{2}\triangle_{Mo(W)}}{|\xi^{e(h)}(\textbf{p}_{e(h)})|^{\frac{3}{2}}}
=\pm\dfrac{\hbar^{3}\alpha_{Mo(W)}^{2}\triangle_{Mo(W)}}{2(\hbar^{2}\triangle_{Mo(W)}^{2}+(\alpha_{Mo(W)} \textbf{p}_{e(h)})^{2})^{\frac{3}{2}}}
\end{equation},

  the  Berry curvature can be then expressed as a function of relative and center of mass momentum (\textbf{p},\textbf{P}) as follows:
 
 \begin{equation} \label{eq:A2}
\boldsymbol{\Omega}_{e(h)}(\textbf{p},\textbf{P})=\pm\dfrac{\hbar^{3}\alpha_{Mo(W)}^{2}\triangle_{Mo(W)}}{2(\hbar^{2}\triangle_{Mo(W)}^{2}+\alpha^{2}_{Mo(W)}(\pm\textbf{p}+ \frac{m_{e(h)}}{M}\textbf{P})^{2})^{\frac{3}{2}}}
\end{equation}.  

$ \boldsymbol{\Omega}_{e(h)}(\textbf{p},\textbf{P}) $ can be treated perturbatively by applying a Taylor expansion    in the second order around \textbf{p} and\textbf{ P}. We obtain then:

\begin{equation}\label{eq:A3}
\boldsymbol{\Omega}_{e(h)}(\pm\textbf{p}+\frac{m_{e(h)}}{M}\textbf{P})=\pm\frac{\alpha^{2}_{Mo(W)}}{2\triangle^{2}_{Mo(W)}}\vec{e}_{z}\mp
\frac{3\alpha^{4}_{Mo(W)}}{4\hbar^{2}\triangle^{4}_{Mo(W)}}(\textbf{p}^{2}+(\frac{m_{e(h)}}{M})^{2}\textbf{P}^{2})
\end{equation}.

Referring to references \citep{zhou, aida, vastana}, the parameter $\alpha^{2}_{Mo(W)}\sim \frac{\hbar^{2}\triangle_{Mo(W)}}{2m_{e(h)}}$. In the limit of zero order, the Berry curvature of charge carriers reaches its maximum, characterized by the following expression:
\begin{equation}\label{eq:A4}
 \boldsymbol{\Omega}_{e(h)}(0,0)=\pm\frac{\hbar^{2}}{4m_{e(h)}\triangle_{Mo(W)}}\vec{e}_{z}. 
 \end{equation}

\end{widetext}


\newpage

\begin{thebibliography}{99}
 

\bibitem{Pettine2023}
J. Pettine, P. Padmanabhan, N. Sirica, R P. Prasankumar, A. J. Taylor and H-T Chen, Ultrafast terahertz emission from emerging symmetry-broken materials. Light Sci. Appl. 12,  133 (2023).

\bibitem{Long2019} 
H. Long, Y. Shi, Qiao Wen and Y H. Tsang, Ultrafast laser pulses (115 fs) generation by using the direct
bandgap ultrasmall 2D GaTe quantum dots, J. Mater. Chem. C. 7, 5937-5944 (2019).

\bibitem{Peres2021}
M. F. C. Martins Quintela, J. C. G. Henriques, and N. M.
R. Peres, Third-order polarizability of interlayer excitons
in heterobilayers, Phys. Rev. B 104, 205433 (2021).

\bibitem{Shree2021}
S. Shree et al, Interlayer exciton mediated second harmonic generation in bilayer MoS2, Nat commun 12, 6894
(2021).

\bibitem{Kim2021}
Y C, Kim, H. Yoo, V T. Nguyen, S. Lee, J-Y Park, and Y H. Ahn, High-Speed Imaging of Second-Harmonic Generation in MoS2
Bilayer under Femtosecond Laser Ablation, Nanomater.11, 1786 (2021).


\bibitem{merkl}
P. Merkl, F. Mooshammer, P. Steinleitner, A. Girnghuber, K-Q. Lin, P. Nagler, J. Holler, C. Schüller, J. M. Lupton, T. Korn, S. Ovesen, S. Brem, E. Malic and R. Hube, Ultrafast transition between exciton phases in van der Waals heterostructures,  Nat. Mat. 18, 691–696 (2019).

\bibitem{donald2017}

F. Wu, T. Lovorn, and A. H. MacDonald, Topological Exciton Bands in Moiré Heterojunctions,  Phys. Rev. Lett .118, 147401
(2017).

\bibitem{Hu2023}
J-X Hu, Y-M Xie, and K. T. Law, Berry curvature, spin Hall effect, and nonlinear optical response in moiré transition metal dichalcogenide heterobilayers, Phys. Rev. B 107, 075424  (2023).
\bibitem{Xie2022}
Y-M Xie, C-P Zhang, J-X Hu, K. F. Mak, and K. Law, Valley-Polarized Quantum Anomalous Hall State in Moiré
$MoTe_{2}/WSe_{2}$ Heterobilayers, Phys. Rev. Lett. 128, 026402 (2022).

\bibitem{Lee2021}
J. Lee, W. Heo, M. Cha, K. Watanabe, T. Taniguchi, J. Kim, S. Cha, D. Kim, M-Ho. Jo, Ultrafast non-excitonic valley
Hall effect in $MoS_{2}/W Te_{2}$ heterobilayers, Nat Comm.12,1635 (2021).

\bibitem{Hu20}
 J-X. Hu, C-P.Zhang, Y-M. Xie and K. T. Law, Nonlinear Hall effects in strained twisted bilayer $WSe_{2}$, Comm Phy.5,255(2022).
 
\bibitem{aida}
A. Hichri, S. Jaziri, M. O. Goerbig, Charged excitons in two-dimensional transition metal dichalcogenides: Semiclassical calculation of Berry curvature effects, Phys. Rev. B. 100, 115426 (2019).

\bibitem{trushin0}
M. Trushin, M. O. Goerbig, and W. Belzig, Optical absorption by Dirac excitons in single-layer transition-metal dichalcogenides, Phys. Rev. B \textbf{94}, 041301(R) (2016).

\bibitem{trushin}
M. Trushin, M. O. Goerbig, and W. Belzig, Model Prediction of Self-Rotating Excitons in Two-Dimensional Transition-Metal Dichalcogenides,  Phys. Rev. Lett. 120,187401 (2018).
\bibitem{vastana}
A. Srivastava and A. Imamoglu, Signatures of Bloch-Band Geometry on Excitons: Nonhydrogenic Spectra in Transition-Metal Dichalcogenides,  Phys. Rev. Lett. 115, 166802
(2015).


\bibitem{zhou}
J. Zhou, W-Y. Shan, W. Yao, and D. Xiao, Berry Phase Modification to the Energy Spectrum of Excitons,  Phys. Rev. Lett. 115,166803 (2015).

\bibitem{Wu2015}
F. Wu, F. Qu, and A. H. MacDonald, Exciton band structure of monolayer $MoS_{2}$,  Phys. Rev. B 91, 075310 (2015).

\bibitem{Chang2021}
Y-W Chang and Y-C Chang, oldy-Wouthuysen transformation for gapped Dirac fermions in two-dimensional semiconducting materials and valley excitons under external fields,  	arXiv:2107.14474 (2021).


\bibitem{Gosselin2006}
P. Gosselin and  F. Ménas and  A. Bérard and  H. Mohrbach, Semiclassical dynamics of electrons in magnetic
Bloch bands: A Hamiltonian approach, Europhys. Lett, 76 (4),  651–656 (2006).

\bibitem{Chang2008}
M-C. Chang and Q. Niu, Berry curvature, orbital moment, and
effective quantum theory of electrons in
electromagnetic fields, J. Phys.: Condens. Matter, 20 193202 (2008).

\bibitem{Gosselin2008}
P. Gosselin and H. Boumrar and H. Mohrbach, Semiclassical quantization of electrons in magnetic fields:
The generalized Peierls substitution, Europhys. Lett, 84 50002 (2008).

\bibitem{Tang2023}
J. Tang, S. Wang and  H. Yu, Inheritance of the exciton geometric structure from Bloch electrons in two-dimensional layered semiconductors, arXiv:2311.04970v1 (2023).

\bibitem{Chaw}
C. K. Yong, M. I. B. Utama, C. S. Ong, T. Cao, E. C. Regan, J. Horng, Y. Shen, H. Cai, K. Watanabe, T. Taniguchi, S. Tongay, H. Deng, A. Zettl, S. G. Louie, F. Wang, Valley-dependent exciton fine structure and Autler–Townes doublets from Berry phases in monolayer $MoSe_{2}$,  Nat Mat.18, 1065–1070 (2019).
\bibitem{tran}
 K. Tran et $al$, Evidence for moiré excitons in van der Waals heterostructures, Nat. 567,71(2019).

\bibitem{evgeny}
E. M. Alexeev et $ al $, Resonantly hybridized excitons in moiré superlattices in van der Waals heterostructures, Nat. Lett. 567, 81–86 (2019).
\bibitem{devakul}
T. Devakul, V. Cr\`epel,Y. Zhang, L. Fu, Magic in twisted transition metal dichalcogenide bilayers, Nat. commun. 12,1(2021).
\bibitem{pasqual}
P. Rivera, H. Yu, K. L. Seyler, N. P. Wilson, W. Yao, X. Xu, Interlayer valley excitons in heterobilayers of transition metal dichalcogenides, 
Nat. Nano. 13,1004–1015 (2018)

\bibitem{donald2018}
F. Wu, T. Lovorn, and A. H. MacDonald, Theory of optical absorption by interlayer excitons in transition metal dichalcogenide heterobilayers,  Phys. Rev. B 97, 035306 (2018).

\bibitem{hichri}
A. Hichri, T. Amand, and S. Jaziri, Resonance energy transfer from moiré-trapped excitons in  
$MoSe_{2}/WSe_{2}$ heterobilayers to graphene: Dielectric environment effect, Phys. Rev. Mat 5, 114002 (2021).

\bibitem{david}
D A. Ruis-Tijerina, I. Soltero and F. Mirels, Theory of moiré localized excitons in transition metal dichalcogenide heterobilayers,  Phys. Rev B 102,195403(2020).

\bibitem{Houssem2023}
H E. Hannachi, D. Elmaghraoui and S. Jaziri. Moiré interlayer exciton relative and center of mass motions coupling. Effect on $1s-np$ interlayer exciton THz transitions, Eur. Phys. J. Plus. 138, 396 (2023).

\bibitem{rivera}
P .Rivera, J. R. Schaibley, A. M. Jones, J. S. Ross $et$ $al$. Observation of long-lived interlayer excitons in monolayer $MoSe_{2}–WSe_{2}$ heterostructures,  Nat. Comm. 6, 6242 (2015).

\bibitem{xiabo}
X. Lu, X. Li and L. Yang. Modulated interlayer exciton properties in a two-dimensional moiré crystal,  Phy. Rev. B. 100, 155416 (2019).


\bibitem{lifetime}
 B. Miller, A. Steinhoff, B. Pano, J. Klein, F. Jahnke, A. Holleitner and U. Wurstbauer, Long-Lived Direct and Indirect Interlayer Excitons in van der Waals Heterostructures, Nano. Lett. 17,9,5229–5237(2017).

\bibitem{Wang2019}
Z. Wang, D. A. Rhodes, K. Watanabe, T. Taniguchi, J. C. Hone, J. Shan, K. F. Mak, Evidence of high-temperature exciton condensation in two-dimensional atomic double layers, Nat. 574, 76–80 (2019).

\bibitem{Gua2022}
H. Guo, X. Zhang and G. Lu, Tuning moiré excitons in Janus heterobilayers for high-temperature Bose-Einstein condensation, Sci. Adv. (2022).
 
\bibitem{Marino}
E. C. Marino et al, Quantum-electrodynamical approach to the exciton spectrum in Transition-Metal
Dichalcogenides, 2D Mater. 5 041006 (2018).



\bibitem{Knapp2021}
S. Chaudhary, C. Knapp, and G. Refael, Anomalous exciton transport in response to a uniform in-plane electric field, Phys. Rev. B 103, 165119 (2021).

\bibitem{Kormanyos2018}
A. Kormányos, V. Zólyomi, V. I. Fal'ko, and G. Burkard, Tunable Berry curvature and valley and spin Hall effect in bilayer $MoS_{2}$, Phys. Rev. B 98, 035408 (2018).

\bibitem{arxiv}
D. Guerci, J. Wang, J. Zang, J.
Cano, J. Pixley, and A. Millis. ArXiv:2207.06476 (2022).



\bibitem{li}
H. Li, S. Li, E. C. Regan, D. Wang, W. Zhao, S. Kahn, K.
Yumigeta, M. Blei, T. Taniguchi, K. Watanabe, Imaging two-dimensional generalized Wigner crystals,
 Nat.597, 650 (2021).

\bibitem{xie}
Y-M. Xie, C-P. Zhang, J-X.
Hu, K. F. Mak, and K. T. Law, Valley-Polarized Quantum Anomalous Hall State in Moiré 
$MoTe_{2}/WSe_{2}$ Heterobilayers, Phys. Rev. Lett.128, 026402 (2022).




\bibitem{keldysh}

L. V. Keldysh, Coulomb interaction in thin semiconductor and semimetal films,  J. Exp. Theor .Phys. 29, 658(1979).

 
\bibitem{robson}

Y. Guo, and J. Robertson, Band engineering in transition metal dichalcogenides: Stacked versus lateral heterostructures,  Appl. Phys. Lett. 108, 233104 (2016).
\bibitem{karplus}

M. Karplus and H. J. Kolker, A Variation‐Perturbation Approach to the Interaction of Radiation with Atoms and Molecules,  J. Chem. Phys 39, 1493 (1963).

\bibitem{xiao}
 D. Xiao, G.-B. Liu, W. Feng, X. Xu, and W. Yao, Coupled Spin and Valley Physics in Monolayers of $MoS_{2}$ and Other Group-VI Dichalcogenides,  Phys. Rev.
Lett. 108,196802 (2012).
\bibitem{tart}
A. Tartakovskii, Excitons in 2D heterostructures,  Nat. Rev. Phys. 2, 8–9 (2020).
\bibitem{gua}
H. Guo, X. Zhang, and G. Lu, Shedding light on moiré excitons: A first-principles perspective,  Sci. Adv. 6, 5638 (2020).
\bibitem{ying}
Y-M Xie, C-P Zhang, J-X Hu, K. F. Mak, and K. Law, Valley-Polarized Quantum Anomalous Hall State in Moiré $MoTe_{2}/WSe_{2}$ Heterobilayers, 
Phys.  Rev. Lett. 128, 026402 (2022).


 \bibitem{dxiao}
 D. Xiao, G.-B. Liu, W. Feng, X. Xu, and W. Yao, Coupled Spin and Valley Physics in Monolayers of 
$MoS_{2}$ and Other Group-VI Dichalcogenides,  Phys. Rev. Lett. 108, 196802 (2012).
 
\bibitem{piechon} 
M. O. Goerbig, G. Montambaux, and F. Piéchon, Measure of Diracness in two-dimensional semiconductors,  Europhys. Lett. 105, 57005 (2014).
\bibitem{simon}
F. Simon, M. Gabay, and M. O. Goerbig and L. Pagot, Role of the Berry curvature on BCS-type superconductivity in two-dimensional materials, 
Phys. Rev. B. 106, 214512 (2022).


\bibitem{bec}
Z. Wang, D. A. Rhodes, K. Watanabe, T. Taniguchi, J. C. Hone, J. Shan, K. F. Mak, Evidence of high-temperature exciton condensation in two-dimensional atomic double layers, Nat.574, 76–80 (2019).



\bibitem{wang}
W. Yao and Q. Niu. Berry Phase Effect on the Exciton Transport and on the Exciton Bose-Einstein Condensate, Phys. Rev. Lett. 101, 106401(2008).
\bibitem{hongli}
H. Guo, X. Zhang and G. Lu, Tuning moiré excitons in Janus heterobilayers for high-temperature Bose-Einstein condensation,  Sci. Adv. (2022).



\bibitem{foldy}
 L. L. Foldy and S. A. Wouthuysen, On the Dirac Theory of Spin 1/2 Particles and Its Non-Relativistic Limit, Phys. Rev. 78, 29 (1950).
 
 
\bibitem{gua}
Y. Guo, J. Robertson, Band engineering in transition metal dichalcogenides: Stacked versus lateral heterostructures,  Appl. Phys. Lett. 108, 233104 (2016).
\end{thebibliography}
\end{document}